\documentclass[12pt]{article}
 \usepackage{graphicx}
 \usepackage[cp1251]{inputenc}

\begin{document}
 \noindent {\footnotesize\it
   Astronomy Letters, 2021, Vol. 47, No 4, pp. 224--234.}
 \newcommand{\dif}{\textrm{d}}

 \noindent
 \begin{tabular}{llllllllllllllllllllllllllllllllllllllllllllll}
 & & & & & & & & & & & & & & & & & & & & & & & & & & & & & & & & & & & & & &\\\hline\hline
 \end{tabular}

  \vskip 0.5cm
\centerline{\bf\large Analysis of Selected Runaway Stars in the Orion Nebula Based}  \centerline{\bf\large on Data from the Gaia EDR3 Catalogue}
   \bigskip
   \bigskip
  \centerline
 {V. V. Bobylev\footnote [1]{e-mail: vbobylev@gaoran.ru}  and A. T. Bajkova}
   \bigskip

  \centerline{\small\it Pulkovo Astronomical Observatory, Russian Academy of Sciences,}

  \centerline{\small\it Pulkovskoe sh. 65, St. Petersburg, 196140 Russia}
 \bigskip
 \bigskip
 \bigskip

{\bf Abstract}---We have performed Monte Carlo simulations of the trajectories of several runaway stars using their parallaxes and proper motions from the Gaia EDR3 catalogue. We have confirmed the hypothesis that the stars AE Aur and $\mu$Col are a product of the multiple system breakup $\sim$2.5 Myr ago and the Orion Trapezium may be the parent cluster for this pair of stars. We show that the data from the Gaia EDR3 catalogue for the star $\iota$Ori, mainly the parallax, do not allow us to talk about the breakup of the multiple
system of AE Aur, $\mu$Col, and $\iota$Ori. The existence of close pair encounters between the stars HD 30112 and HD 43112 $\sim$1 Myr ago has been confirmed. Close triple encounters confirm the hypothesis that the stars HD 30112 and HD 43112 escaped from the parent cluster Col 69. We show that the stars HIP 28133 and TYC 5368-1541-1 have a nonzero probability of escape from the region within 10 pc of the center of the Orion Trapezium cluster and a fairly high probability (about 8\%) that they were both at distances less than 20 pc from the center of the Orion Trapezium $\sim$2.5 Myr ago. It has been established for the
first time that the stars Gaia EDR3 3021115184676332288 and Gaia EDR3 2983790269606043648 have a probability of about 0.5\% that they broke up as a binary system $\sim$1.1 Myr ago. The star Gaia EDR3 3021115184676332288 has a probability of about 16\% that it escaped from the region within 10 pc of the center of the Orion Trapezium cluster $\sim$1 Myr ago.


 \subsection*{INTRODUCTION}
It was shown in the pioneering paper by Blaauw and Morgan (1954) that two stars, AE Aur and $\mu$Col, move apart in opposite directions with high velocities, more than 100~km s$^{-1}$, while their trajectories extended into the past intersect $\sim$2.6 Myr ago in the
Orion Nebula. The paper by Blaauw and Morgan aroused great interest of researchers in questions related to the detection and study of young runaway stars not only in the Orion association, but also in other OB associations located near the Sun (Hoogerwerf et al. 2001; Tetzlaff et al. 2011; Maiz-Apell\'aniz et al. 2018). Dozens of stars running radially away
from the center of the Orion Nebula with high velocities have been detected to date (McBride and Kunkel 2019; Platais et al. 2020; Farias et al. 2020; Schoettler et al. 2020).

Analysis of the data from the Hipparcos (1997) catalogue led Hoogerwerf et al. (2001) to conclude that the observed properties of three runaway stars, AE Aur, $\mu$Col, and $\iota$Ori, are consistent with their common origin $\sim$2.5 Myr ago. The Monte Carlo
simulations performed by these authors showed that the parent cluster for these three stars is closely associated with the Orion Trapezium. They also pointed out that the most probable mechanism producing the high velocities of the stars is the collision and disruption
of two binaries. This scenario was proposed by Gies and Bolton (1986).

Several more remarkable runaways are known in the Orion Nebula. For example, Hoogerwerf
et al. (2001) showed that the stars HD 30112 and HD 43112 were ejected from the compact open star cluster Col 69 ($\lambda$~Ori). The Kleinmann–Low (KL) nebula containing a source of radio and infrared emission known as the Becklin-Neugebauer (BN) object is located at the center of the Orion Nebula. Here, several sources are seen to move apart in a region called Orion BN/KL. In particular, Rodriguez et al. (2017) found that the disintegration of a multiple stellar system occurred in the 15th century or, more specifically, in the year $1475\pm6.$

The Orion Trapezium is a massive compact open star cluster located right at the center of the Orion Nebula, in the Orion belt. This cluster contains a number of very young massive stars. The brightest components of $\theta^1$~Ori (A,B,C,D, and E) form a figure in the shape of a trapezium in the visible sky. This multiple system is known to be gravitationally
unstable (Olivares et al. 2013; Allen et al. 2017).

AE~Aur ($V=5^m.96$) and $\mu$~Col ($V=5^m.18$) are single young massive stars, both of spectral type O9.5V. The star $\iota$Ori ($V=2^m.77$) is known to be a multiple system consisting of four components (Maiz-Apell\'aniz and Barb\'a 2020). The primary component Aa is a spectroscopic binary consisting of O9III and B1III stars (Marchenko et al. 2000), the component Ab is a B2IV star, and the most distant component B of the system is a B2V star.

The accuracy of stellar parallaxes and proper motions is of paramount importance for the simulations of spatial stellar trajectories. At present, the Gaia EDR3 (Gaia Early Data Release 3, Brown et al. 2020; Lindegren et al. 2020) version has appeared, where, in comparison with the previous Gaia DR2 version (Brown et al. 2018), the trigonometric parallaxes and proper motions were improved approximately by 30\% for $\sim$1.5 billion stars.

The goal of this paper is to analyze the trajectories of several runaway stars using up-to-date data from the Gaia EDR3 catalogue. We want to find out how close the encounters between these stars were and how close the encounters of the stars with the center of the putative parent open cluster could be. To solve this problem, we chose two young open star clusters in Orion, the Orion Trapezium and Col 69.

 \subsection*{DATA}
{\bf AE Aur, $\mu$Col, and $\iota$Ori.} Table 1 gives the measured characteristics of the stars AE Aur (HD 34078, Hip 24575), $\mu$Col (HD 38666, Hip 27204), and $\iota$Ori
(HD 37043, Hip 26241) taken from several sources. These include, first, the Hipparcos (1997) catalogue with which Hoogerwerf et al. (2001) worked, second, the Gaia DR2 catalogue (Brown et al. 2018) that was used by Schoettler et al. (2020) in searching for runaway stars, third, the Gaia EDR3 catalogue, and, finally, the VLBI measurements for the Orion Nebula that have a high accuracy of measuring the trigonometric parallaxes and proper motions of radio stars and masers.

We see that there are no significant differences in the initial data for the star AE Aur. The heliocentric distance $d=1/\pi$ calculated for the star $\mu$Col based on data from the Gaia EDR3 catalogue exceeds considerably the one found from the Hipparcos (1997) catalogue. In addition, it can be seen that the random errors in the parameters of the stars AE Aur and $\mu$Col in the Gaia EDR3 catalogue are approximately half those in the Gaia DR2 catalogue and a factor of 5--10 smaller than those in the Hipparcos (1997) catalogue. It is important to note that for the stars AE Aur and $\mu$Col the relative errors of the parallaxes (and,
accordingly, the distances) given in the last column of Table 1 do not exceed 5\%. Therefore, for these stars there is no need to take into account the Lutz–Kelker (1973) bias, which depends strongly on the parallax errors.

In contrast, the proper motions for the star $\iota$Ori from the Hipparcos (1997) and Gaia EDR3 catalogues differ greatly, with the random errors of the parameters for this star being smaller in the Hipparcos (1997) catalogue. Thus, new simulations are of great interest. In this paper for this star we use its systemic heliocentric radial velocity from Marchenko et al. (2000).

Finally, Table 1 gives the proper motions and trigonometric parallaxes measured by the VLBI
method. We took these data from Reid et al. (2019). Here, the radio observations of several radio stars and masers observed in the Orion Nebula were averaged. These were mostly T Tauri stars. These data refer to the center of the Orion Nebula and the Orion Trapezium, in particular. The trigonometric parallax (and, hence, the distance) was determined here with
a very high accuracy. We took the heliocentric radial velocity from Kharchenko et al. (2005), which was calculated from the Orion Trapezium stars. Note that Schoettler et al. (2020) used slightly different proper motions and radial velocity for the center of the
Orion Nebula, $\mu_\alpha\cos\delta=1.51\pm0.11$~mas yr$^{-1}$, $\mu_\delta=0.50\pm0.12$~mas yr$^{-1}$ and $V_r=21.8\pm6.6$~km s$^{-1}$, which they took from Kuhn et al. (2019). The proper motions and radial velocities for the Orion Trapezium used by us are such that a wide range of values, into which the values from Kuhn et al. (2019) also fall, is covered in our Monte Carlo simulations.

Using data from the GaiaDR2 catalogue, Schoettler et al. (2020) and Farias et al. (2020) selected dozens of suitable candidates for runaway stars that could escape from the Orion Nebula in the past. From these publications we selected such stars that, first, possessed a significant (more than 35 km s$^{-1}$) escape velocity from the Orion Nebula, second, had measured radial velocities, and, third, had measurements in the Gaia EDR3 catalogue. The most interesting stars from the selected ones are given in Table 2. These stars were used in this paper, in particular, in searching for a possible member of the multiple system containing AE Aur and $\mu$Col, possibly $\iota$Ori, and possibly yet another member.

 \begin{table}[t]                               
 \caption[]{\small\baselineskip=1.0ex\protect
Initial kinematic characteristics of the stars taken from various catalogues, $\mu_\alpha^*=\mu_\alpha\cos\delta$
 }
 \begin{center}
  \begin{tabular}{|r|c|c|c|c|c|c|c|}\hline
 \label{tab-dat-1}
 \def\baselinestretch{1}\normalsize\small
                          Star & Hipparcos (1997) & Gaia DR2 & Gaia EDR3 & VLBI \\\hline
                          AE Aur & &  & & \\
 $\mu_\alpha^*,$ mas/yr & $-4.05\pm0.66$ & $-4.440\pm0.115$ & $-4.747\pm0.046$ & \\
   $\mu_\delta,$ mas/yr & $43.22\pm0.44$ & $43.368\pm0.081$ & $43.538\pm0.033$ & \\
          $\pi,$ mas    & $~2.24\pm0.74$ & $~2.464\pm0.066$ & $~2.574\pm0.034$ & \\
          $V_r,$ km/s   & $ 57.5\pm1.2 $ (a) &        & $  56.7\pm0.6 $ (b) &\\
            $d,$ pc     & $446^{+220}_{-111}$ &     & $388^{+6}_{-5}$ &\\
                  \hline
                       $\mu$ Col & &  & & \\
 $\mu_\alpha^*,$ mas/yr & $~3.01\pm0.52$ & $~2.988\pm0.279$ & $~3.271\pm0.095$ & \\
           $\mu_\delta,$ mas/yr &$-22.62\pm0.50$ &$-22.030\pm0.291$ &$-22.176\pm0.110$ & \\
                  $\pi,$ mas    & $~2.52\pm0.55$ & $~2.148\pm0.162$ & $~1.702\pm0.090$ & \\
                  $V_r,$ km/s   & $109.0\pm2.5 $ (c) &          & $ 109.0\pm1.8 $ (d) &\\
                    $d,$ pc     & $397^{+110}_{-~70}$ &     & $588^{+32}_{-30}$ &\\
                  \hline
                     $\iota$~Ori & &  & & \\
 $\mu_\alpha^*,$ mas/yr  & $~2.27\pm0.65$ & & $-2.816\pm1.022$ & \\
   $\mu_\delta,$ mas/yr  & $-0.62\pm0.47$ & & $-1.693\pm0.833$ & \\
          $\pi,$ mas     & $~2.46\pm0.77$ & & $~1.997\pm0.730$ & \\
          $V_r,$ km/s    & $ 28.7\pm1.1 $  (e) & & $  31.3\pm1.2$ (f) &\\
            $d,$ pc      & $406^{+185}_{-~96}$ & & $501^{+288}_{-134}$ &\\
                  \hline
  Orion  & & & & \\
  Trapezium & & & & \\
 $\mu_\alpha^*,$ mas/yr  & & & & $~~3.14\pm3.06 $ \\
   $\mu_\delta,$ mas/yr  & & & & $~-1.19\pm3.71 $ \\
          $\pi,$ mas     & & & & $~2.421\pm0.019$ \\
          $V_r,$ km/s    & & & & $~~28.9\pm2.7~ $ (g) \\
            $d,$ pc      & & & & $413^{+3}_{-4}$ \\
                  \hline
 \end{tabular} \end{center}
{\small The radial velocities were taken from the following papers: (a), (c), and (d)—Hoogerwerf et al. (2001), (b)—Kharchenko et al. (2007),
(d)—Gontcharov (2006), (f)—Marchenko et al. (2000), (g)—Kharchenko et al. (2005).
}  \end{table}
 \begin{table}[t]
 \caption[]{\small\baselineskip=1.0ex\protect
 Initial kinematic characteristics taken from the Gaia EDR3 catalogue for runaway stars from the lists by Schoettler et al. (2020) and Farias et al. (2020)
 }
 \begin{center}
 \begin{tabular}{|r|r|r|r|r|r|r|r|}\hline
 \label{tab-dat-2}

 \def\baselinestretch{1}\normalsize\small
  Gaia EDR3 &  $\pi$      & $\mu_\alpha\cos\delta$ & $\mu_\delta$ & $V_r$ \\
            &     mas     &    mas/yr  &   mas/yr  & km/s  \\\hline
3012438796685305728&$2.403\pm0.016$ & $~20.116\pm0.015$ &$-21.574\pm0.014$ & $~~0.0\pm6.8$\\
2986587942582891264&$2.616\pm0.039$ & $-16.632\pm0.037$ &$-22.102\pm0.035$ & $-21.3\pm6.6$\\
2989899774685582592&$2.199\pm0.014$ & $-11.062\pm0.013$ &$-12.579\pm0.011$ & $~~3.5\pm6.6$\\
2998537847270106240&$2.307\pm0.017$ & $~10.916\pm0.015$ &$-10.408\pm0.014$ & $~45.3\pm6.6$\\
3003060825792025088&$2.425\pm0.015$ & $~15.327\pm0.017$ &$~-4.445\pm0.015$ & $~36.9\pm6.7$\\
3021115184676332288&$2.510\pm0.013$ & $~29.568\pm0.015$ &$~~2.294\pm0.013$ & $~31.9\pm0.8$\\
2983790269606043648&$2.752\pm0.012$ & $~-3.901\pm0.010$ &$-34.407\pm0.011$ & $~16.6\pm1.4$\\
   \hline
   \end{tabular} \end{center}
 \end{table}
 \begin{table}[t]
 \caption[]{\small\baselineskip=1.0ex\protect
 Initial kinematic characteristics of the stars associated with the cluster Col 69 taken from the Gaia EDR3 catalogue
 }
 \begin{center}
 \begin{tabular}{|r|r|r|r|r|r|r|r|}\hline
 \label{tab-Col69}

 \def\baselinestretch{1}\normalsize\small
  Object &  $\pi$      & $\mu_\alpha\cos\delta$ & $\mu_\delta$ & $V_r$ & Ref $V_r$\\
         &     mas     &    mas/yr  &   mas/yr  & km/s & \\\hline

HD 43112 & $2.566\pm0.082$ & $~25.786\pm0.087$ &$~11.591\pm0.070$ & $36.9\pm0.7$ & (a) \\
HD 30112 & $2.765\pm0.047$ & $-45.744\pm0.051$ &$-31.284\pm0.041$ & $~9.0\pm4.4$ & (a) \\
  Col 69 & $2.551\pm0.090$ & $~~0.741\pm0.236$ &$~-2.015\pm0.165$ & $27.5\pm0.4$ & (b) \\
   \hline
   \end{tabular} \end{center}
{\small (a) Gontcharov (2006) and (b) Carrera et al. (2019). }
 \end{table}

{\bf HD 30112, HD 43112, and Col 69.} The runaway stars HD~30112 ($V=7^m.22$) and HD~43112 ($V=5^m.89$) are young massive stars of spectral types B3/5V and B1V, respectively. The open star cluster Col 69 is quite rich in stars. According to Cantat-Gaudin et al. (2018), this cluster includes 669 stars from the Gaia DR2 catalogue.

For the stars HD 30112 (HIP 22061), HD 43112 (HIP 29678), and the cluster Col 69 Table 3 gives their trigonometric parallaxes and proper motions taken from the Gaia EDR3 catalogue. To calculate the mean trigonometric parallax and proper motions of the cluster Col 69, we took the measurements from the Gaia EDR3 catalogue. For this purpose, from the catalogue by Cantat-Gaudin et al. (2018) we selected several tens of stars with a cluster membership
probability greater than 0.9 within 7' of the cluster center that have numbers according to the Gaia DR2 catalogue.

 \subsection*{METHOD}
The axisymmetric Galactic potential $\Phi$ in which the stellar trajectories are calculated into the past is represented as a sum of three components~--- a central spherical bulge $\Phi_b$, a disk $\Phi_d$, and a massive spherical dark matter halo $\Phi_h$:
 \begin{equation}
 \begin{array}{lll}
  \Phi=\Phi_b+\Phi_d+\Phi_h.
 \label{pot}
 \end{array}
 \end{equation}
The potentials of the bulge $\Phi_b$ and the disk $\Phi_d$ are represented in the form proposed by Miyamoto and Nagai (1975), while the halo component is represented according to Navarro et al. (1997). The specific values of the parameters of this model Galactic potential are given in Bajkova and Bobylev et al. (2016), where it is designated as model III.

In the heliocentric coordinate system the $x$ axis is directed toward the Galactic center, the $y$ axis is in the direction of Galactic rotation, and the $z$ axis is directed to the north Galactic pole. Then, $x=d\cos l\cos b,$ $y=d\sin l\cos b$ and $z=d\sin b,$ where
$d=1/\pi$ is the stellar heliocentric distance in kpc that we calculate via the stellar parallax $\pi$ in mas.

The Galactic orbits of the stars are constructed in the coordinate system associated with the local standard of rest. Therefore, the peculiar velocity of the Sun with $(U,V,W)_\odot=(11.0,12.0,7.2)$~km s$^{-1}$ from Sch\"onrich et al. (2010) is excluded from the initial data. We also take into account the height of the Sun above the Galactic plane $h_\odot=16$~pc (Bobylev and Bajkova 2016).

For each pair of stars we calculate the parameter of the encounter between their orbits $\Delta d(t)=\sqrt{\Delta x^2(t)+\Delta y^2(t)+\Delta z^2(t)}$. Then, we determine the parameters $\Delta d_{\rm min}$ at the encounter time $t_{\rm min}$. We estimate the errors in $\Delta d_{\rm min}$ and $t_{\rm min}$ by the Monte Carlo method. The errors of the stellar parameters are assumed to be distributed normally with an rms deviation $\sigma$. The errors are added to the stellar proper motion components, parallax, and radial velocity.

 \begin{figure} [t] {\begin{center}
  \includegraphics[width=150mm]{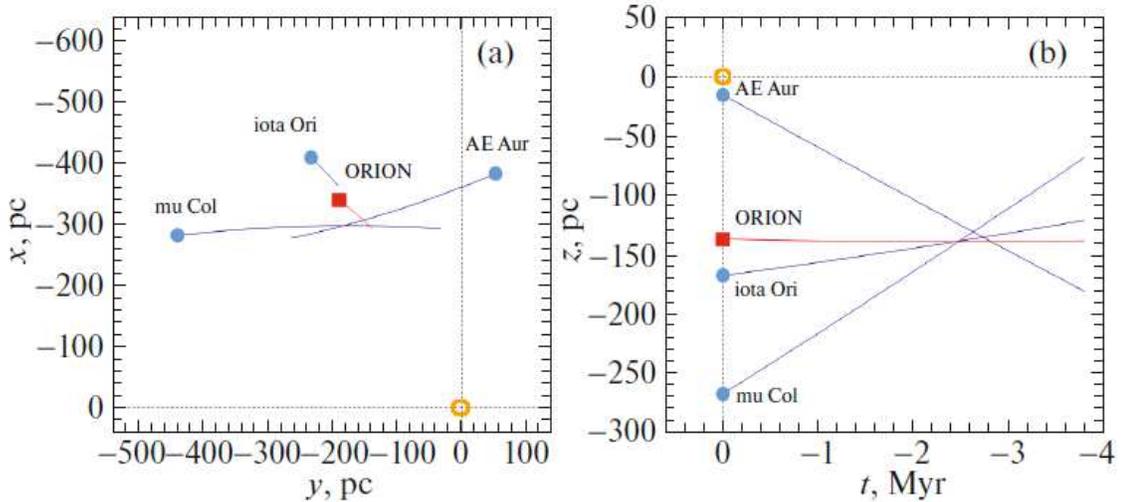}
 \caption{
(a) The stellar trajectories in projection onto the Galactic $xy$ plane constructed back into the past and (b) coordinate $z$ versus integration time $t$; the current positions of the stars are marked by the blue circles, the red color indicates the position
and trajectory of the Orion Nebula, the yellow ring marks the Sun's position.
 }
 \label{f-1}
 \end{center} }
 \end{figure}
 \begin{figure} [t] {\begin{center}
  \includegraphics[width=150mm]{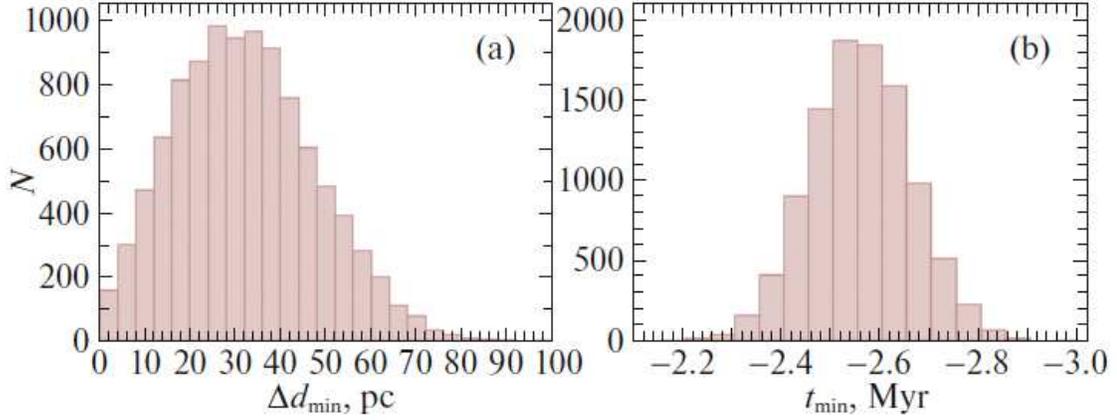}
 \caption{Histograms of the minimum separation $\Delta d_{min}$~(a) and encounter times $t_{min}$~(b) between the two stars AE Aur and $\mu$~Col.
 }
 \label{f-2}
 \end{center} }
 \end{figure}
 \begin{figure} [t] {\begin{center}
  \includegraphics[width=150mm]{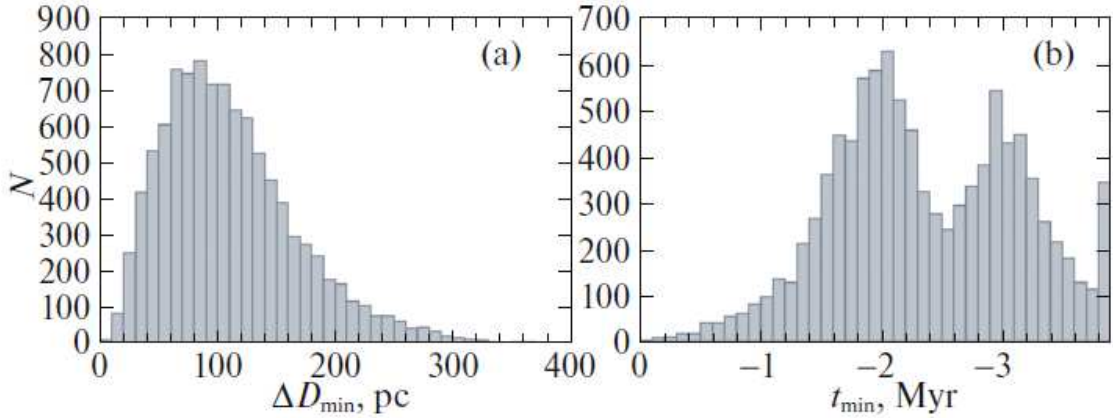}
 \caption{Histograms of the minimum separation $\Delta D_{min}$~(a) and encounter times $t_{min}$~(b) between the three stars AE Aur, $\mu$~Col and $\iota$~Ori.
 }
 \label{f-3}
 \end{center} }
 \end{figure}
 \begin{figure} [t] {\begin{center}
  \includegraphics[width=150mm]{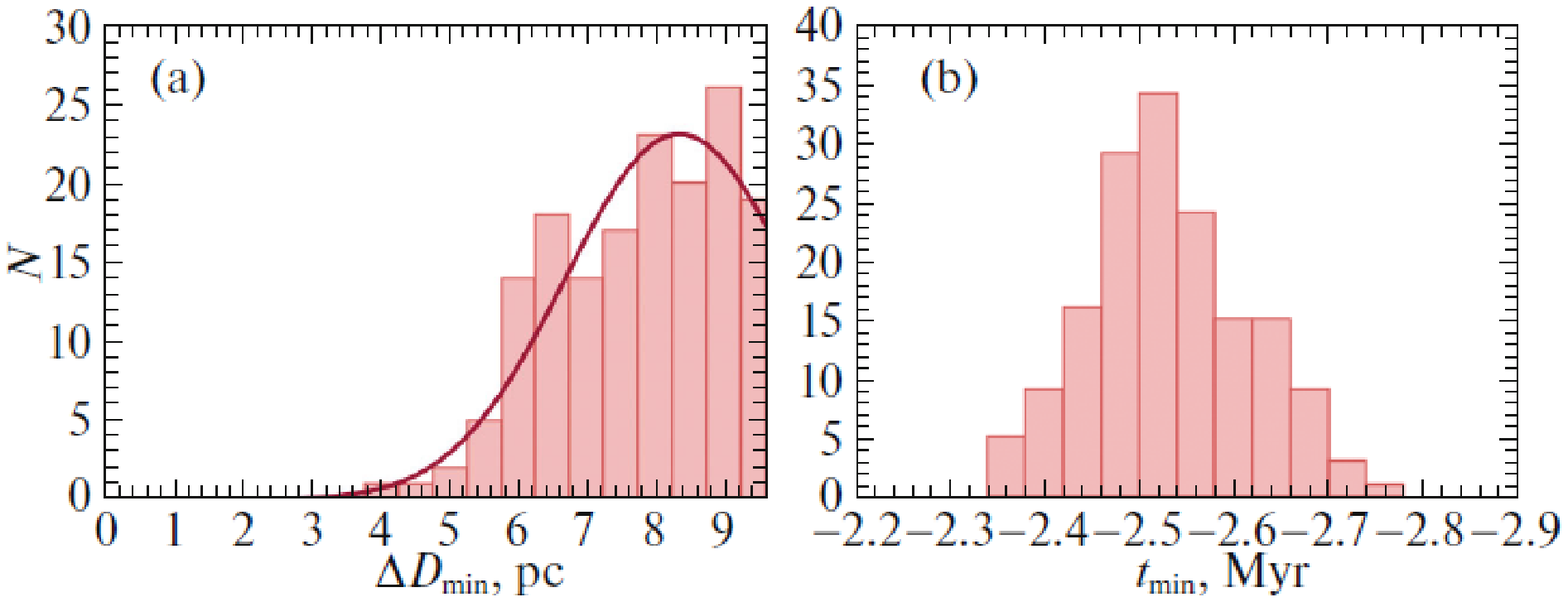}
\caption{Histograms of the minimum separation $\Delta D_{min}$~(a) and times of triple encounters $t_{min}$~(b) between the star AE Aur, $\mu$~Col, and the Orion Trapezium selected under the condition $\Delta D_{min}<10$~pc.
 }
 \label{f-4}
 \end{center} }
 \end{figure}

When considering triple encounters, following Hoogerwerf et al. (2001), we define the separation between the three stellar systems $\Delta D_{\rm min} (t)$ as the maximum deviation of the objects from their average position. Thus, $\Delta D_{\rm min}= max|\textbf{x}_j-{\overline \textbf{x}}|$, where, for example, $j=$AE Aur, $\mu$Col, and $\iota$Ori, ${\overline \textbf{x}}=(\textbf{x}_{AE Aur}+\textbf{x}_{\mu Col}+\textbf{x}_{\iota Ori})/3$, is the average position, and $\textbf{x}_j$ is the position of star $j.$

The distribution of the triple encounter parameter in some neighborhood can be depicted as a histogram. We calculate the expected distribution $F_{3D}$ of the minimum separation $\Delta D_{min}$ using the formula from Hoogerwerf et al. (2001)
 \begin{equation}
\renewcommand{\arraystretch}{1.8}
 \begin{array}{lll} \displaystyle
   F_{3D}(\Delta D_{min})= \frac{ \Delta D_{min} }{2\sigma\mu\sqrt\pi}
 \times\biggl\{\exp\biggl[-\frac{(\Delta D_{min}-\mu)^2}{4\sigma^2}\biggr]
              -\exp\biggl[-\frac{(\Delta D_{min}+\mu)^2}{4\sigma^2}\biggr]\biggr\}
 \label{F3D}
 \end{array}
 \end{equation}
for the appropriate mean $\mu$ and standard deviation $\sigma$.

The size of the region around the open star cluster where the star remains gravitationally bound varies greatly, depending on the cluster mass and stellar density distribution. The tidal radii may be used. Piskunov et al. (2008) determined them for more than 600 open star clusters. For example, for Col 69 ($\sim10^2~M_\odot$) the tidal radius was found to be 7.7~pc. For the more massive ($\sim10^3~M_\odot$) Orion Trapezium cluster it may be smaller, because most of its mass is known to be concentrated within $<$3~pc (Da Rio et al. 2014). Nevertheless, in our Monte Carlo simulations we take a typical value of 10 pc as the tidal
radius of the open star cluster.

Both the Orion Trapezium and Col 69 are very young open star clusters. According to Piskunov
et al. (2008), the age of Col 69 found from isochrones is $\sim$5.7 Myr ($\log t=6.76$). Da Rio et al. (2014) estimated the formation time of the Orion Nebula to be $\sim$4 Myr. Hence, we integrate the orbits of the stars under study for 5--6 Myr into the past.

 \begin{figure} [t] {\begin{center}
  \includegraphics[width=150mm]{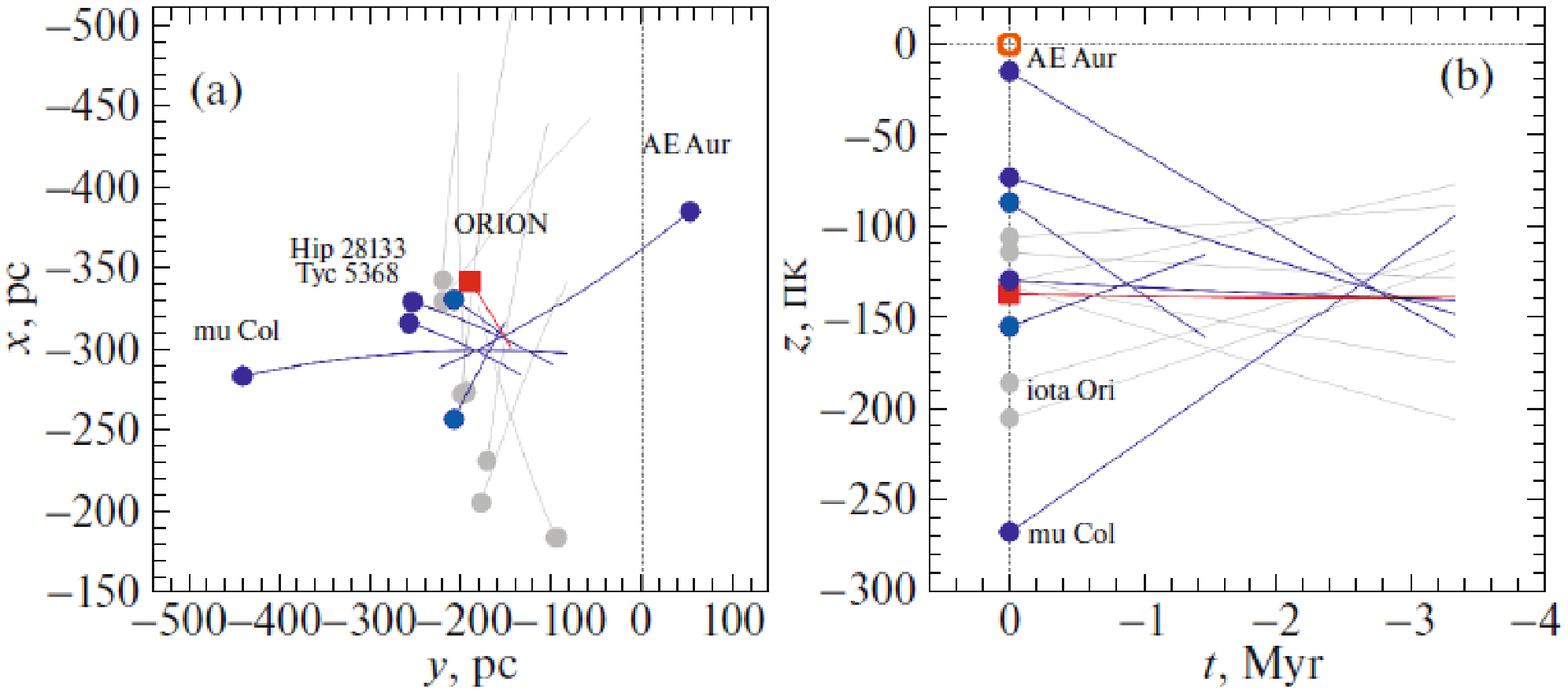}
 \caption{
(a) The stellar trajectories in projection onto the Galactic $xy$ plane constructed back into the past and (b) coordinate $z$ versus integration time $t$; the current positions of the stars are marked by the circles, the red color indicates the position and trajectory of the center of the Orion Trapezium cluster; for details, see the text.
 }
 \label{f-22-poisk}
 \end{center} }
 \end{figure}
 \begin{figure} [t] {\begin{center}
 \includegraphics[width=150mm]{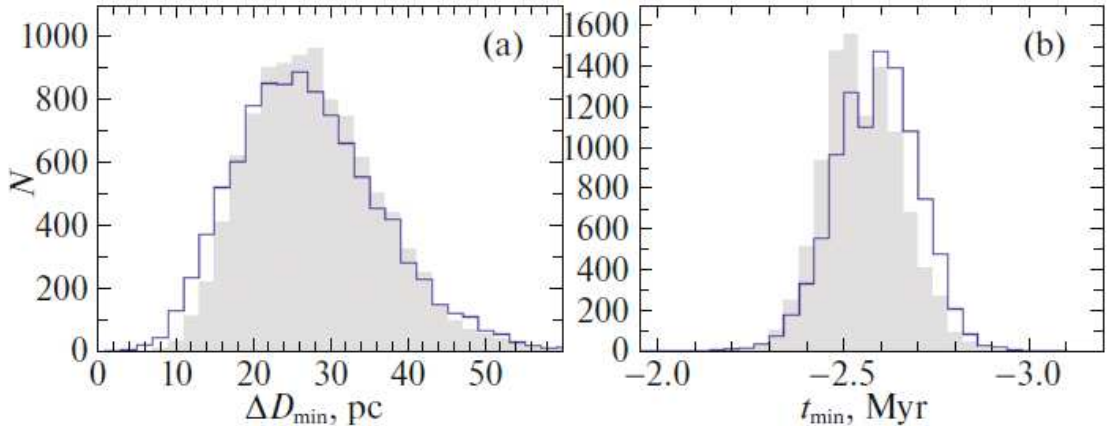}
 \caption{
Histograms of the minimum separation $\Delta D_{min}$~(a) and times of triple encounters $t_{min}$~(b) between the stars AE Aur, $\mu$~Col, and each of the stars HIP 28133 (gray shading) or TYC 5368-1541-1 (dark contour).
 }
 \label{f-troynoe-Hip}
 \end{center} }
 \end{figure}
 \begin{figure} [t] {\begin{center}
  \includegraphics[width=150mm]{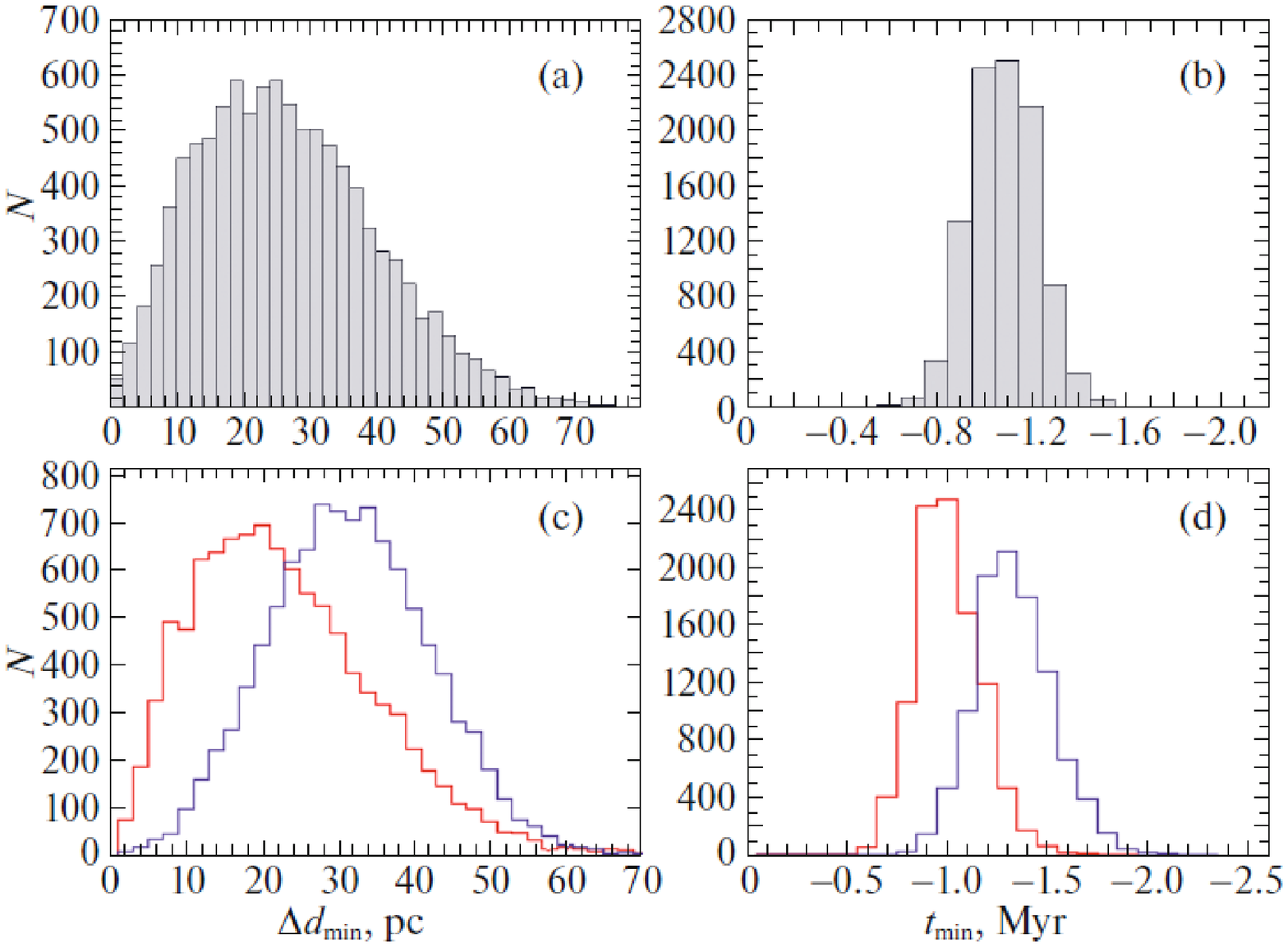}
 \caption{
Histograms of the minimum separation $\Delta d_{min}$~(a) and times of pair encounters $t_{min}$~(b) between the two starsGaiaEDR3 3021115184676332288 and Gaia EDR3 2983790269606043648; histograms of the minimum separation $\Delta d_{min}$~(c) and times of pair encounters $t_{min}$~(d) between the center of the Orion Nebula and each of the stars Gaia EDR3 3021115184676332288 (red contour) or Gaia EDR3 2983790269606043648 (dark contour).
 }
 \label{f-parnoe-EDR3}
 \end{center} }
 \end{figure}

 \subsection*{RESULTS AND DISCUSSION}
{\bf AE Aur, $\mu$Col, and $\iota$Ori.} Figure 1a presents the trajectories of the stars AE Aur, $\mu$Col, $\iota$Ori, and the Orion Trapezium in projection onto the Galactic $xy$ plane, while Fig. 1b depicts their vertical motion in the past on an integration interval $\sim$4~Myr.

Figure 2 presents the histograms of the minimum pair separation $\Delta d_{min}$ and encounter times $t_{min}$ between two stars, AE Aur, $\mu$Col, and $\iota$Ori. To construct the model orbits of these stars, we made 10 000 realizations by the Monte Carlo method.

As can be seen from Figs. 1 and 2, the trajectories of the stars AE Aur and $\mu$Col intersect. As a result of our detailed analysis of 10 000 model data, we found that there are very close encounters: 15 and 31 encounters within 1 and 2 pc, respectively. We
can conclude that our simulations confirm the hypothesis by Blaauw and Morgan (1954) that the stars AE Aur and $\mu$Col are a binary disruption product. The disruption event occurred $\sim$2.5 Myr ago. This conclusion is also consistent with the results of an analysis of the orbits for these two stars obtained by Hoogerwerf et al. (2001) based on data from the
Hipparcos (1997) catalogue.

A completely different situation is observed with regard to the triple encounter of AE Aur, $\mu$Col, and $\iota$~Ori. Figure 3 presents the histograms of the minimum separation $\Delta D_{min}$ and encounter times $t_{min}$ between them. To construct the model orbits of these stars, we made 10 000 realizations by the Monte Carlo method. The maximum of the distribution of the minimum separation $\Delta D_{min}$ lies near 80 pc (Fig. 3a), which is in conflict with the results of other authors. For example, when simulating similar encounters using data from the Hipparcos (1997) catalogue, Hoogerwerf et al. (2001) constructed (Fig. 8 in their paper) the histogram of $\Delta D_{min}$ with its maximum near 6 pc, while the standard deviation was about 2 pc. These authors also noted that out of the 2.5 million model orbits of the three stars, very close encounters, where $\Delta D_{min}<1$~pc, were detected in 114 cases.

The unusual double-humped distribution in Fig. 3b is explained by the fact that the star $\iota$Ori approaches the trajectories of the other two stars at different times. Thus, the arrival of the star $\iota$Ori from far away is reflected here.

The star $\iota$Ori is very bright, $V=2^m.77,$ which cannot but affect the quality of its measurements. The parallax of the star $\iota$Ori was apparently measured more reliably in the Hipparcos catalogue, because the telescope of this program was oriented to the measurements of considerably brighter stars compared to the Gaia satellite.

Figure 4 presents the histograms of the minimum separation $\Delta D_{min}$~(a) and times of triple encounters $t_{min}$~(b) between the stars AE Aur, $\mu$Col, and the Orion Trapezium selected under the condition $\Delta D_{min}<10$~pc. The exponential in Fig. 4a is fitted according to Eq. (2) with the mean $\mu=8$~pc and rms deviation $\sigma=1.2$~pc. Thus, out of the 10 000 model encounters, we obtained 160 in which all three objects are within 10 pc. This result confirms the hypothesis that the Orion Trapezium may be the parent cluster for the pair of AE Aur and $\mu$Col.

The distribution of encounter times $t_{min}$ in Fig. 4b has a maximum near 2.5 Myr in the past, in good agreement with the times found for the pair encounters of AE Aur and $\mu$Col (Fig. 2b).

Note that the star $\iota$Ori and Orion lie on the same line of sight, as can be clearly seen from Fig. 1a. Thus, if we took the parallax of the star $\iota$Ori from the Hipparcos (1997) catalogue (see Table 1), then we would get close triple encounters.

 \begin{figure} [t] {\begin{center}
  \includegraphics[width=150mm]{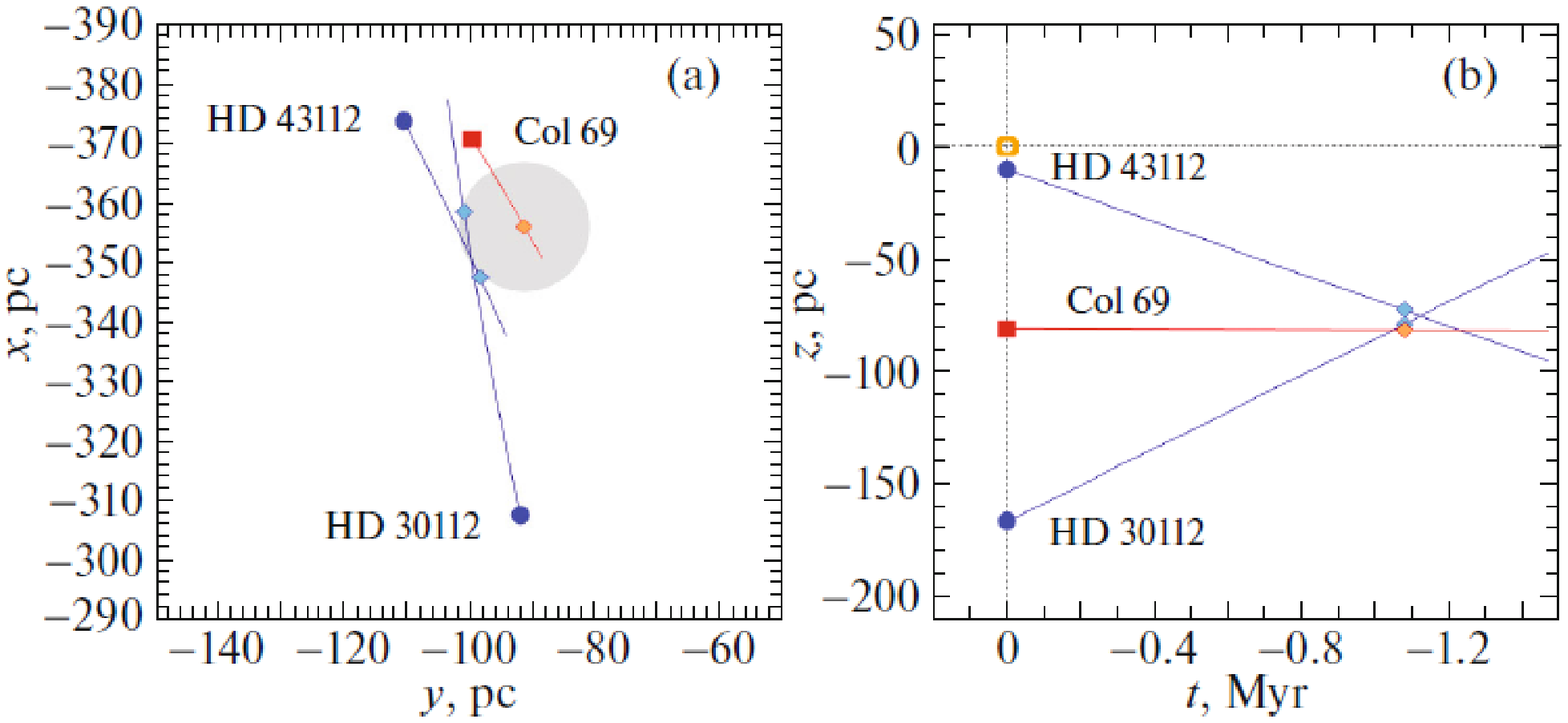}
 \caption{
(a) The trajectories of the stars HD 30112, HD 43112, and the cluster Col 69 in projection onto the Galactic $xy$ plane constructed back into the past and (b) coordinate $z$ versus integration time $t$; the current positions of the stars are marked by the blue circles, the red color indicates the position and trajectory of the center of the cluster Col~69, the diamonds mark the positions of the objects at the time of their closest encounter, the gray circle indicates the neighborhood of radius 10~pc with the center in the cluster Col 69 at the time of the closest encounter with the stars, the yellow ring marks the position of the Sun.
 }
 \label{f-5col69}
 \end{center} }
 \end{figure}

{\bf HIP 28133 and TYC 5368-1541-1.} Figure 5 presents the projections of the spatial trajectories constructed for the stars AE Aur, $\mu$Col, the Orion Trapezium, and the stars from Table 2. The gray color marks the positions and trajectories that are not of
great interest for the solution of our problems. There are even more such stars than is given in Table 2, because it does not include the stars with parallaxes greater than 3~mas that are available in the list by Farias et al. (2020), but they did not enter into the
figure. The darker color in the figure marks the most interesting stars.

These primarily include the two stars HIP 28133 (Gaia EDR3 2998537847270106240) and TYC5368-1541-1 (Gaia EDR3 3003060825792025088) with some of their characteristics given in the fourth and fifths rows of Table 2. These stars were selected as candidates for runaway stars by Schoettler et al. (2020). HIP 28133 is known to be a star of spectral type G8IV. As can be seen from Fig. 5, the projections of their orbits intersect with the projections
of the orbits of the stars AE Aur and $\mu$Col.

Hence, we carried out an experiment to find out what probability that one of these stars or both of them could be a member of a disrupted multiple system in the past is. The result is reflected in Fig. 6. It can be concluded that all these stars were at an average distance from one another of about 25 pc $\sim$2.5 Myr ago. Thus, they were in the same parent cluster, but there were no very close (within 1--2 pc) encounters both between the triplet of HIP 28133, AE Aur, and $\mu$Col and the triplet of TYC 5368-1541-1, AE Aur, and $\mu$Col.

We also found that the stars HIP 28133 and TYC 5368-1541-1 have a high probability that they
escaped from the region within 10 pc of the center of the Orion Trapezium $\sim$2.5 Myr ago. For example, for the star HIP 28133 we found 92 events at a distance to the Orion Trapezium $\Delta d_{min}<2$~pc and 2406 events at $d_{min}<10$~pc (24\%). For the star TYC 5368-1541-1 this number is more modest --- 831 events at a distance to the center of the Orion Trapezium $\Delta d_{min}<10$~pc (8

When studying the pair encounters of these stars, we found that $\sim$2 Myr ago they were spaced less than 2 and 4 pc apart in 16 and 69 cases, respectively, although the maximum of the corresponding distribution occurs at $\Delta d_{min}\sim35$~pc. Thus, although the
binary breakup scenario involving these two stars has a moderately high probability, it must not be ruled out.

{\bf Gaia EDR3 3021115184676332288 and Gaia EDR3 2983790269606043648.}
The characteristics of these stars are given in the last two rows of Table 2. These are stars from the list by Farias et al. (2020). In Fig. 5 the positions of these stars are marked by the blue circles. As can be seen from Fig. 5b, the projections of their orbits onto the $t-z$ plane intersect $\sim$1 Myr ago and then they diverge dramatically. Therefore, they are shown shorter than the remaining projections.

Our analysis of the pair encounters showed that these stars have a fairly high probability that $\sim$1.1 Myr ago they broke up as a binary system. For example, at $\Delta d_{min}<2$~pc we found 49 events (0.5\%). The histograms of pair encounters between these two stars are presented in Figs. 7a and 7b. Comparing these results, for example, with the distributions of pair encounters between the stars AE Aur and $\mu$Col in Fig. 2 (the maximum of the distribution is located near 30 pc in Fig. 2a), we see that in Fig. 7a (the
maximum near 20 pc) the encounters are closer. Thus, the hypothesis about the binary breakup in this case is better supported by our simulations.

Both these stars are quite faint; their observations are available only in the Gaia DR2 and Gaia EDR3 catalogues. Both have $G\sim12^m$; so far there is no other information about them.

Figures 7c and 7d present the histograms of the minimum separation $\Delta d_{min}$ and times of pair encounters $t_{min}$ between the center of the Orion Nebula and each of the stars Gaia EDR3 3021115184676332288 or Gaia EDR3 2983790269606043648. It can be seen that the star Gaia EDR3 3021115184676332288 was fairly close ($\Delta d_{min}<10$~pc) to the center of the Orion Nebula $\sim$1 Myr ago. Here, the probability of close encounters is about 16\%.

 \begin{figure} [t] {\begin{center}
  \includegraphics[width=140mm]{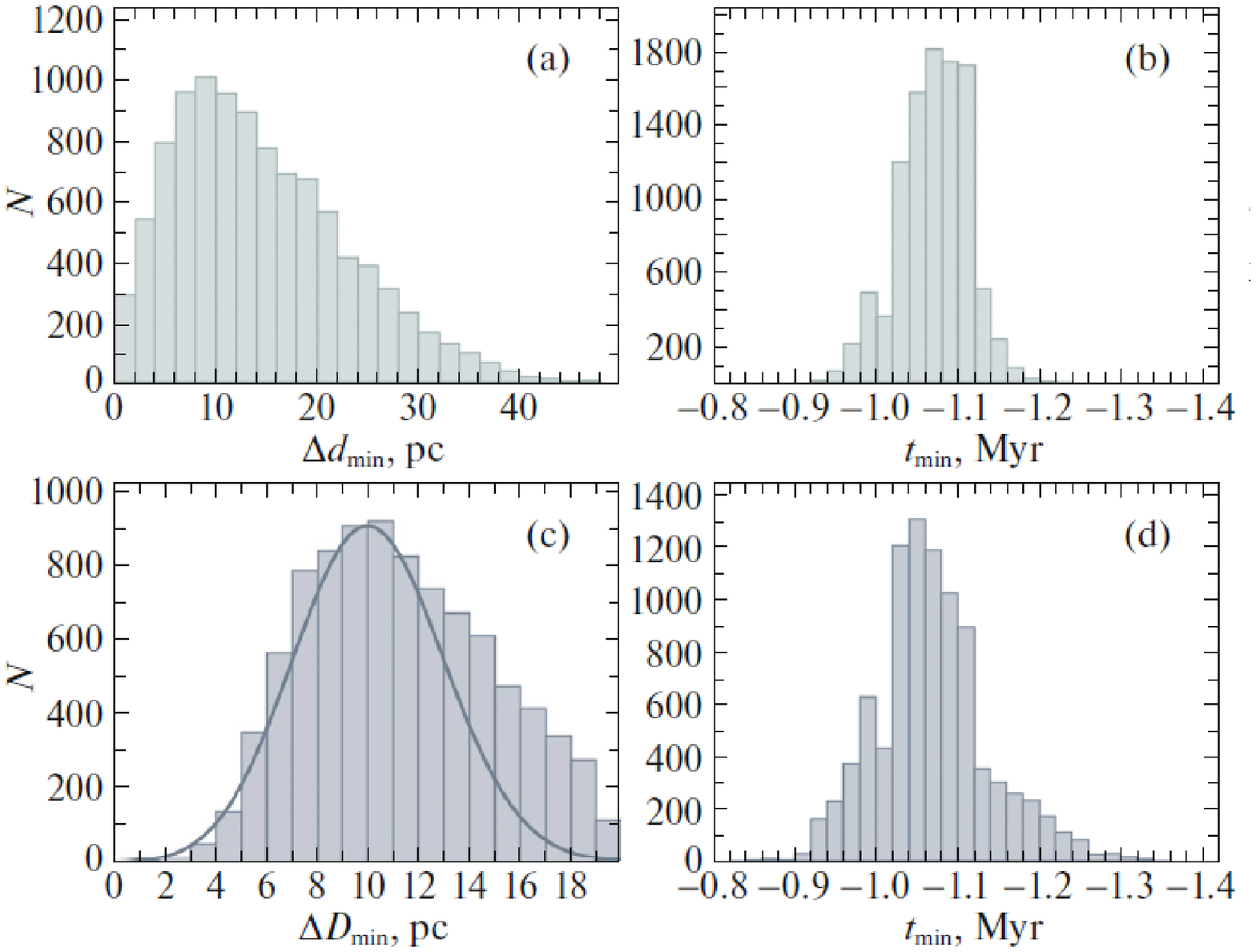}
 \caption{
Histograms of the minimum separation for a pair encounter $\Delta d_{min}$~(a) and encounter times $t_{min}$~(b) between the stars HD 30112 and HD 43112, the minimum separation for a triple encounter $\Delta D_{min}$~(c), and encounter times $t_{min}$~(d) between the
stars and Col 69 selected under the condition?Dmin $\Delta D_{min}<20$~pc.
 }
 \label{f-6col69}
 \end{center} }
 \end{figure}

{\bf HD 30112, HD 43112, and Col 69.} Figure 8 presents the spatial trajectories of the starsHD30112, HD43112, and Col 69 constructed back into the past. These trajectories were constructed using nominal kinematic data. We can see that at $t=-1.06$~Myr the relative separation between the stars HD 30112 and HD 43112 did not exceed 10 pc, and they were both
within 10 pc of the center of the cluster Col 69. Our Monte Carlo simulations of encounters including the errors in the measurements show that the encounters between the objects under consideration could be even closer. To construct the model orbits of the stars, we made 10 000 realizations by the Monte Carlo method. The results of our simulations are presented in Fig. 9.

Figures 9a and 9b present the histograms of the minimum separation $\Delta d_{min}$~(a) and times of pair encounters $t_{min}$ between the stars HD 30112 and HD 43112. Very close encounters are found here--- 35 events at $\Delta d_{min}<1$~pc and 148 events at $\Delta d_{min}<2$~pc (the event probability is 1.5\%).

Using data from the Hipparcos catalogue (van Leeuwen 2007), Bobylev and Bajkova (2009)
found the probability of pair encounters within $\Delta d_{min}\leq10$~pc to be 6\% when simulating this pair of stars. According to the histogram in Fig. 9a, the probability
of pair encounters between the stars HD 30112 and HD 43112 within $\Delta d_{min}\leq10$~pc is 11\%. Note that in this paper we used the data in which the errors in the proper motions and parallaxes of the stars HD 30112 and HD 43112 are smaller than those for these stars
in the van Leeuwen (2007) catalogue by an order of magnitude.

Figures 9c and 9d present the histograms of the minimum separation $\Delta D_{min}$~(a) and times of triple encounters $t_{min}$ between the stars HD 30112, HD 43112, and the center of the cluster Col 69 selected under the condition $\Delta D_{min}<20$~pc. The exponential in Fig. 9c is fitted according to Eq. (2) with the mean $\mu=9$~pc and rms deviation $\sigma=2.2$~pc. We see that there are very close triple encounters, for example, 10 events at $\Delta D_{min}<4$~pc. In addition, at $\Delta D_{min}<10$~pc there are 3569 triple encounters, accounting for 36\% of the 10 000 model realizations. Thus, our simulations confirm the hypothesis that the cluster Col 69 ($\lambda$Ori) could be the parent cluster for the pair of HD 30112 and HD 43112, which broke up 1.1 Myr ago within this cluster.

 \subsection*{CONCLUSIONS}
We performed Monte Carlo simulations of the trajectories of several runaway stars using their parallaxes and proper motions taken from the Gaia EDR3 catalogue.

First we considered the well-known pair of the stars AE Aur and $\mu$Col and simulated both their pair and triple encounters involving the Orion Trapezium cluster. We showed that the stars AE Aur and $\mu$Col could be a product of the multiple system breakup $\sim$2.5 Myr ago, while the Orion Trapezium could be the parent cluster for this pair of stars. Thus, the
hypothesis by Blaauw and Morgan (1954) about a possible place of escape of these runaway stars was confirmed.

The data from the Gaia EDR3 catalogue for the star $\iota$Ori, mainly the parallax, do not allow us to talk about a common origin of the stars AE Aur, $\mu$Col, and $\iota$Ori, i.e., according to the Gaia EDR3 data, $\iota$Ori is too far away from the Sun and far from the place of possible disruption of the multiple system of AE Aur and $\mu$Col in the past. The star $\iota$Ori is very bright, $G=2.74^m.$ Its parallax may have been measured better in the Hipparcos catalogue. The conclusions reached by Hoogerwerf et al. (2001) about a common
origin of the multiple system of AE Aur, $\mu$Col, and $\iota$Ori then remain valid.

We considered the pair of HD~30112 and HD~43112
that are associated with the cluster Col 69. The
existence of close pair encounters between these stars
1.1 Myr ago was confirmed. Close triple encounters
confirm the hypothesis that the stars HD~30112 and
HD~43112 escaped from the parent cluster Col~69
1.1 Myr ago. Here, the errors in the proper motions
and parallaxes of the stars under study from the
Gaia EDR3 catalogue turned out to be smaller than
those used before by an order of magnitude, while
the probabilities of these hypothesis turned out to be
twice their previous estimates obtained in a similar
way.

Two results related to our analysis of the motion of the runaway stars in the Orion Trapezium are completely new.

First, we showed that the stars HIP 28133 (Gaia EDR3 2998537847270106240) and TYC5368-1541-1 (Gaia EDR3 3003060825792025088) have a nonzero probability of escape from the region within 10 pc of the Orion Trapezium center and a probability of about 8\% that they were both at distances less than 20~pc from the Orion Trapezium center $\sim$2.5 Myr ago. Having analyzed the pair encounters of these stars, we found a nonzero probability that they were
formed by a binary breakup $\sim$2 Myr ago.

Second, we showed that the stars Gaia EDR3 3021115184676332288 and Gaia EDR3~2983790269606043648 have a nonzero probability that they broke up as a binary system $\sim$1.1 Myr ago. For example, we found 49 model events (out of 10 000 realizations) in which the separation between these two stars was less than 2~pc. The star Gaia EDR3
3021115184676332288 has a probability of 16\% that it escaped from the Orion Trapezium, because it was close ($\Delta d_{min}<10$~pc) to the Trapezium center $\sim$1 Myr ago.

\bigskip{\bf ACKNOWLEDGMENTS}

We are grateful to the referee for the useful remarks that contributed to an improvement of the paper.

\bigskip \bigskip\medskip{\bf REFERENCES}{\small

1. C. Allen, R. Costero, A. Ruelas-Mayorga, and L. J. S\'anchez, Mon. Not. R. Astron. Soc. 466, 4937 (2017).

2. A. T. Bajkova and V. V. Bobylev, Astron. Lett. 42, 567 (2016).

3. A. Blaauw and W. W. Morgan, Astrophys. J. 119, 625 (1954).

4. V. V. Bobylev and A. T. Bajkova, Astron. Lett. 35, 396 (2009).

5. V. V. Bobylev and A. T. Bajkova, Astron. Lett. 42, 1 (2016).

6. A. G. A. Brown, A. Vallenari, T. Prusti, J. H. J. de Bruijne, C. Babusiaux, C. A. L. Bailer-Jones, M. Biermann, D. W. Evans, et al. (Gaia Collab.), Astron. Astrophys. 616, 1 (2018).

7. A. G. A. Brown, A. Vallenari, T. Prusti, J. H. J. de Bruijne, C. Babusiaux, M. Biermann,
O. L. Creevey, D. W. Evans, et al. (Gaia Collab.), arXiv: 2012.01533 (2020).

8. T. Cantat-Gaudin, C. Jordi, A. Vallenari, A. Bragaglia, L. Balaguer-Nu\~nez, C. Soubiran, D. Bossini, A. Moitinho, et al., Astron. Astrophys. 618, 93 (2018).

9. R. Carrera, A. Bragaglia, T. Cantat-Gaudin, A. Vallenari, L. Balaguer-Nu\~nez, D. Bossini, L. Casamiquela, C. Jordi, et al., Astron. Astrophys. 623, 80 (2019).

10. J. P. Farias, J. C. Tan, and L. Eyer, Astrophys. J. 900, 14 (2020).

11. D. R. Gies and C. T. Bolton, Astrophys. J. Suppl. 61, 419 (1986).

12. G. A. Gontcharov, Astron. Lett. 32, 759 (2006).

13. The HIPPARCOS and Tycho Catalogues, ESA SP--1200 (1997).

14. R. Hoogerwerf, J. H. J. de Bruijne, and P. T. de Zeeuw, Astron. Astrophys. 365, 49 (2001).

15. N. V. Kharchenko, A. E. Piskunov, S. R\"oser, E. Schilbach, and R.-D. Scholz, Astron. Astrophys. 438, 1163 (2005).

16. N. V. Kharchenko, R.-D. Scholz, A. E. Piskunov, S. R\"oser, and E. Schilbach, Astron. Nachr. 328, 889 (2007).

17. M. A. Kuhn, L. A. Hillenbrand, A. Sills, E. D. Feigelson, and K. V. Getman, Astrophys. J. 870, 32 (2019).

18. F. van Leeuwen, Astron. Astrophys. 474, 653 (2007).

19. L. Lindegren, S. A. Klioner, J. Hern\'andez, A. Bombrun, M. Ramos-Lerate, H. Steidelm\"uller, U. Bastian, M. Biermann, et al. (Gaia Collab.), arXiv:
2012.03380 (2020).

20. T. E. Lutz and D. H. Kelker, Publ. Astron. Soc.Pacif. 85, 573 (1973).

21. J. Maiz-Apell\'aniz, M. P. Gonz\'alez, R. H. Barb\'a, S. Sim\'on-Diaz, I. Negueruela, D. J. Lennon, A. Sota, and E. Trigueros P\'aez, Astron. Astrophys. 616, 149 (2018).

22. J. Maiz-Apell\'aniz and R. H. Barb\'a, Astron. Astrophys. 636, 28 (2020).

23. S. V. Marchenko, G. Rauw, E. A. Antokhina, I. I. Antokhin, D. Ballereau, J. Chauville, M.~F.~Corcoran, R. Costero, et al., Mon. Not. R. Astron. Soc. 317, 333 (2000).

24. A. McBride and M. Kunkel, Astrophys. J. 884, 6 (2019).

25. M. Miyamoto and R. Nagai, Publ. Astron. Soc. Pacif. 27, 533 (1975).

26. J. F. Navarro, C. S. Frenk, and S. D. M. White, Astrophys. J. 490, 493 (1997).

27. J. Olivares, L. J. Sanchez, A. Ruelas-Mayorga, C. Allen, R. Costero, and A. Poveda, Astron. J. 146, 106 (2013).

28. A. E. Piskunov, E. Schilbach, N. V. Kharchenko, S. R\"oser and R.-D. Scholz, Astron. Astrophys. 477, 165 (2008).

29. I. Platais, M. Robberto, A. Bellini, V. Kozhurina-Platais, M. Gennaro, G. Strampelli, L. A. Hillenbrand, S. E. de Mink, et al., Astron. J. 159, 272 (2020).

30. M. J. Reid, K. M. Menten, A. Brunthaler, X. W. Zheng, T. M. Dame, Y. Xu, J. Li, N. Sakai, et al., Astrophys. J. 885, 131 (2019).

31. N. da Rio, J. C. Tan, and K. Jaehnig, Astrophys. J. 795, 55 (2014).

32. L. F. Rodriguez, S. A. Dzib, L. Loinard, L. Zapata, L. G\'omez, K. M. Menten, and S. Lizano, Astrophys. J. 834, 140 (2017).

33. C. Schoettler, J. de Bruijne, E. Vaher, and R. J. Parker, Mon. Not. R. Astron. Soc. 495, 3104 (2020).

34. R. Sch\"onrich, J. Binney and W. Dehnen, Mon. Not. R. Astron. Soc. 403, 1829 (2010).

35. N. Tetzlaff, R. Neuh\"auser, and M. M. Hohle, Mon. Not. R. Astron. Soc. 410, 190 (2011).  }
  \end{document}